\documentstyle[12pt,epsfig,amssymb]{article}
\begin{document}
\setlength{\parskip}{0.45cm}
\setlength{\baselineskip}{0.75cm}
\begin{titlepage}
\begin{flushright}
DO-TH 98/04 \\ 
DTP/98/22 \\
April 1998 \\
\end{flushright}
\vspace{0.1cm}
\begin{center}
\Large
{\bf {Next-to-Leading Order QCD Corrections to the}}

\vspace{0.1cm}
{\bf {Polarized Photoproduction of Heavy Flavors}}

\vspace{1.2cm}
\large
I.\ Bojak\\

\vspace*{0.5cm}
\normalsize
{\it Institut f\"{u}r Physik, Universit\"{a}t Dortmund, D-44221 Dortmund,
Germany}

\vspace*{0.3cm}
and\\  

\vspace*{0.6cm}
\large
M.\ Stratmann\\

\vspace*{0.5cm}
\normalsize
{\it Department of Physics, University of Durham, Durham DH1 3LE, England}

\vspace{0.9cm}
%
\large
{\bf Abstract} \\
\end{center}
\vspace{0.3cm}
We present a calculation of the next-to-leading order 
$({\cal{O}}(\alpha_s^2 \alpha))$ QCD corrections to heavy flavor
photoproduction with longitudinally polarized beams. We apply
our results to study the longitudinal spin asymmetry for the
total charm quark production cross section which will be utilized by the 
forthcoming COMPASS experiment at CERN to obtain first direct
information on the polarized gluon density $\Delta g$. 
We also briefly discuss the main theoretical uncertainties 
inherent in this calculation. In particular we demonstrate that 
the factorization scale dependence is considerably 
reduced in next-to-leading order.
\end{titlepage}
\newpage
%
%
\noindent
Despite significant experimental progress in spin-dependent
deep inelastic scattering (DIS) \cite{exp}, the polarized parton densities
$\Delta f$ $(f=q,\,\bar{q},\,g)$ still remain considerably less known 
than their unpolarized counterparts $f$.
In particular the polarized gluon density $\Delta g$
is almost completely unconstrained in all recent
next-to-leading order (NLO) analyses \cite{grsv,gs,pdfs} of presently
available DIS data.
This is due to the limited kinematical coverage of the data which does
not allow for an unambiguous determination of $\Delta g$ from scaling
violations. An even more important role plays the lack of a momentum
sum rule and the absence of any direct constraints from other processes 
like jet or direct photon production, both being available and
important for the determination of unpolarized parton densities.
Therefore a series of `next generation' spin experiments will put a special
emphasis on such exclusive measurements to provide further invaluable 
information for a more restrictive analysis of polarized parton
distributions in the future.

The first direct information on $\Delta g$ is expected to be provided
by the COMPASS fixed target experiment at CERN \cite{compass}
from studies of the longitudinal spin asymmetry of total open charm
photoproduction. Heavy quark $(Q=c,\,b)$ production is generally 
considered to be one of the best options to pin down $\Delta g$
since in leading order (LO) only the photon-gluon fusion (PGF) process 
\cite{lohq}
\begin{equation}
\label{eq:lopgf}
\vec{\gamma} \vec{g} \rightarrow Q \bar{Q}
\end{equation}
contributes (an arrow denotes a longitudinally polarized particle).
All studies of this process so far \cite{compass,lohq,svhera,other} 
were limited to LO estimates which, however, 
are known to be notoriously unreliable due to, e.g., their strong 
factorization/renormalization scale dependence.
In addition, the already available unpolarized NLO corrections 
\cite{svn,ellis} turn out to be sizeable in certain kinematical regions 
and the `clean picture' of Eq.~(\ref{eq:lopgf}) is obscured 
by genuine NLO subprocesses with light quarks in the initial state. 
The knowledge of the polarized NLO cross section is thus mandatory 
for a meaningful extraction of $\Delta g$.
It is the main purpose of this paper to provide for the first time 
the results of such a complete NLO $({\cal{O}}(\alpha_s^2 \alpha))$
QCD calculation. Thereby we hope to provide a more reliable
theoretical basis for the measurement of $\Delta g$ by the COMPASS 
collaboration.
 
The remainder of the paper is organized as follows: next we briefly sketch
the most important details of our calculation, mainly concentrating on
complications which arise due to the presence of polarized
particles in the initial state. We then apply our results to present for
the first time estimates for the longitudinal spin asymmetry of total
open charm photoproduction in NLO QCD in the kinematical region 
accessible by the COMPASS experiment \cite{compass}. 
Finally, we address
the relevance of the main theoretical uncertainties inherent
in this calculation. For more details on our calculation, analytical
NLO results, and studies of $p_T$ and rapidity differential heavy 
quark cross sections, we refer the interested reader to 
Ref.\ \cite{longpaper}.

%

The NLO QCD corrections to the PGF mechanism in Eq.~(\ref{eq:lopgf})
consist of three parts:
\begin{description}
\item[(i)] the one-loop virtual corrections,
\item[(ii)] the real corrections with an additional gluon in the final state
\begin{equation}
\label{eq:nlopgf}
\vec{\gamma} \vec{g} \rightarrow Q \bar{Q} g \quad ,
\end{equation}
\item[(iii)] a new production mechanism appearing for the 
first time in NLO
\begin{equation}
\label{eq:nloqua}
\vec{\gamma} \vec{q}\; (\vec{\bar{q}}) \rightarrow Q \bar{Q} q\; (\bar{q})
\quad .
\end{equation}
\end{description}
As is well known one encounters various types of singularities when
calculating the loop- and $2\rightarrow 3$ phase space integrals.
Ultraviolet (UV) singularities which show up only in the virtual
corrections (i) are removed by on-shell mass and wavefunction and
(modified) $\overline{\mathrm{MS}}$ coupling constant renormalization. 
In the latter case one also removes the gluon self-energy contribution 
due to the heavy quark loop in addition to the subtractions 
usually performed in the $\overline{\mathrm{MS}}$ scheme 
(see, e.g., \cite{svn,longpaper,nason}). This explicit decoupling
of the heavy quark at low energy leads to a fixed flavor
scheme with $n_{lf}$ light flavors (active
in the running of $\alpha_s$ and in the parton evolution) and
the produced heavy flavor. 
Infrared (IR) divergencies and double pole terms appearing in the
non-abelian parts when IR and mass/collinear (M) singularities
coincide cancel in the sum of (i) and (ii).   
Finally one is left with the M divergencies in (ii) and (iii) which
are removed by the factorization procedure in the 
$\overline{\mathrm{MS}}$ scheme. 

One has to choose a consistent method of regularizing these
singularities so that they become manifest. 
For this purpose we choose to work in the well-established framework of 
$n$-dimensional regularization. 
When calculating the required helicity dependent matrix elements 
we have to project onto the helicity states of the incoming particles. 
This is achieved by using the standard relations (see, e.g., \cite{craigie})
\begin{equation}
\label{eq:polgluon}
\epsilon_{\mu}(k_1,\lambda_1)\, \epsilon^*_{\nu}(k_1,\lambda_1) =
\frac{1}{2} \left[-g_{\mu\nu} + i \lambda_1 \epsilon_{\mu\nu\rho\sigma}
\frac{k_1^{\rho} k_2^{\sigma}}{k_1 \cdot k_2} \right]
\end{equation}
for incoming photons with momentum $k_1$ and helicity $\lambda_1$
(accordingly for gluons with $k_2$ and $\lambda_2$) and
\begin{equation}
\label{eq:polquark}
u(k_2,h) \bar{u}(k_2,h) = \frac{1}{2} \not\! k_2 (1-h \gamma_5)
\end{equation}
for incoming quarks with momentum $k_2$ and helicity $h$
(analogously for antiquarks).
Using (\ref{eq:polgluon}) and (\ref{eq:polquark}) we can calculate the 
contributions to heavy flavor photoproduction with unpolarized 
{\em{and}} polarized beams {\em{simultaneously}} by taking the sum
or the difference of the helicity dependent squared matrix elements
\begin{eqnarray}
\label{eq:unpme}
\mathrm{unpolarized:} \quad \overline{\left| M \right|}^{\: 2} &=&
\frac{1}{2} \left[ \left|M\right|^2(++) + \left|M\right|^2(+-)\right]\\
\label{eq:polme}
\mathrm{polarized:} \quad  \Delta\left|M\right|^2 &=&
\frac{1}{2} \left[ \left|M\right|^2(++) - \left|M\right|^2(+-)\right]
\end{eqnarray}
where $\left|M\right|^2(h_1,h_2)$ denotes the squared matrix element for
any of the contributing subprocesses (\ref{eq:lopgf}) and (i)-(iii)
for definite helicities $h_1$ and $h_2$ of the two incoming particles.
The possibility to obtain the unpolarized 
results\footnote{Since each
boson has $n-2$ degrees of freedom, one has to average the
spin of each incoming photon (gluon) with a factor $1/(n-2)$ replacing
the $1/2$ in (\ref{eq:polgluon}) for the unpolarized
case. Furthermore, the inclusion of external ghost contributions (see, e.g.,
\cite{longpaper}) allows us to drop all terms other than
$g_{\mu\nu}$ in the symmetric part of (\ref{eq:polgluon}).} `for free'
provides an important check on the correctness of our results. We
fully agree with the unpolarized results 
presented in \cite{svn} and \cite{ellis}.
The abelian part of our result agrees analytically with the 
results of two recent calculations of the NLO QCD corrections to
$\vec{\gamma}\vec{\gamma}\rightarrow Q\bar{Q}$ \cite{conto} as well.

The presence of $\gamma_5$ and the totally anti-symmetric tensor
$\epsilon_{\mu\nu\rho\sigma}$ in the polarized calculation due to
(\ref{eq:polgluon}) and (\ref{eq:polquark}) introduces
some complications because their
purely four-dimensional origin allows no straightforward continuation to
$n\neq 4$ dimensions. We choose to work in the HVBM scheme \cite{hvbm}
which was shown to provide an internally consistent extension of
$\gamma_5$ and $\epsilon_{\mu\nu\rho\sigma}$ to arbitrary
dimensions\footnote{This prescription is incorporated in {\tt TRACER}
\cite{tracer}, which we have used for all trace calculations.}.
The price to pay is that apart from the usual
$n$-dimensional scalar products $k\cdot p$ their respective 
$(n-4)$-dimensional subspace counterparts, usually 
denoted by $\widehat{k\cdot p}$ (`hat momenta'), can also
show up in $\left|M\right|^2(h_1,h_2)$ in the HVBM scheme. For the 
single-inclusive heavy quark production considered here,
one can choose a convenient frame (see \cite{longpaper} for
details) where all non-vanishing $(n-4)$ scalar products can be 
expressed by a single hat momenta combination $\hat{p}^2=-\widehat{p\cdot p}$.
The latter quantity appears exclusively in the polarized 
$2\rightarrow 3$ contributions and originates from scalar products of the 
momenta of the two not observed (integrated) final state particles
in (\ref{eq:nlopgf}).
However, terms proportional to $\hat{p}^2$ still deserve special attention
when performing the $2\rightarrow 3$ phase space integrations since 
the $(n-4)$-dimensional subspace cannot be trivially integrated out as in an
unpolarized calculation.

The appropriately modified phase space formula for the present
polarized calculation can be schematically expressed as follows
\begin{equation}
\label{eq:dps3}
\mathrm{dPS}_3 = \mathrm{dPS}_{3,unp}(\theta_1,\theta_2) \times
\frac{1}{B\left(\frac{1}{2},\frac{n-4}{2}\right)} \int_0^1 dx\,
\frac{x^{(n-6)/2}}{\sqrt{1-x}}
\end{equation}
with
\begin{equation}
\label{eq:x}
x\equiv \hat{p}^2/\hat{p}^2_{\mathrm{max}} = \frac{4 (s_4+m^2) \hat{p}^2}
{s_4^2 \sin^2\theta_1 \sin^2\theta_2}
\end{equation}
where $B$ is the Euler Beta function, $m$ denotes the mass of the heavy
quark,  and 
$s_4\equiv s+t_1+u_1$ is the sum of the three $2\rightarrow 2$ Mandelstam
variables. The angles $\theta_{1,2}$ are introduced to parametrize the momenta
of the two integrated partons (see \cite{longpaper}). 
Since
\begin{equation}
\label{eq:hats}
\frac{1}{B\left(\frac{1}{2},\frac{n-4}{2}\right)} \int_0^1 dx\,
\frac{x^{(n-6)/2}}{\sqrt{1-x}}\left\{\begin{array}{c}1\\
\hat{p}^2\end{array}\right. =\left\{\begin{array}{c}1\\ 
(n-4)\, s_4^2 \sin^2\theta_1 \sin^2\theta_2 /[4 (s_4+m^2)]\end{array}\right. ,
\end{equation}
Eq.~(\ref{eq:dps3}) reduces
to the well-known `unpolarized' phase space $\mathrm{dPS_{3,unp}}$
(see, e.g., \cite{smith2})
for the vast majority of terms without a $\hat{p}^2$ dependence and
for the rest, which is proportional to $\hat{p}^2$,
introduces a $n \rightarrow n+2$ shift in the angular integrals
(\ref{eq:integrals}) discussed below, guaranteeing collinear safety.
Due to the extra powers of $s_4$ all these additional integrals
are infrared safe as well,
implying that {\em{all}} contributions due to $\hat{p}^2$ 
are at least of ${\cal{O}}(n-4)$ (cf.\ Eq.~(\ref{eq:hats})) 
and hence drop out when the limit $n \rightarrow 4$ is taken\footnote{This 
differs from a calculation \cite{werner} involving only {\em massless}
particles, where IR poles can occur.}. 

The remaining phase space integration then proceeds as in the 
unpolarized case \cite{svn}. One has to make
extensive use of relations between Mandelstam variables
to reduce complex combinations to simpler ones by partial
fractioning. In the end only integrals containing
at most two angular dependent Mandelstam variables remain 
\begin{equation}
\label{eq:integrals}
I^{(k,l)} = \int_0^{\pi} d\theta_1 \int_0^{\pi} d\theta_2
\frac{\sin^{n-3}\theta_1
\sin^{n-4}\theta_2}{\left(a+b \cos \theta_1\right)^k
\left(A+B \cos\theta_1+C \sin\theta_1 \cos\theta_2\right)^l}
\end{equation}
where $a$, $b$, $A$, $B$, and $C$ are only functions of $s$, $t_1$,
$u_1$, and $m^2$. A useful list of integrals of the type 
(\ref{eq:integrals}) can be found in \cite{smith2}, but we have
checked all required formulae.
The loop integrations have been performed
using a Passarino-Veltman reduction \cite{pass} to scalar
integrals which can also be found in \cite{smith2} and we have recalculated 
them as well.

In the soft gluon limit for (\ref{eq:nlopgf}) the $2\rightarrow 3$ 
kinematics reduces to the usual $2\rightarrow 2$ Born kinematics 
and soft gluon poles associated with $1/s_4$ terms occur, 
since $s_4\rightarrow 0$ when the final state gluon momentum $k_3$ 
becomes soft $(k_3\rightarrow 0)$. To deal with these poles one can 
simply slice the phase space into a soft gluon
$(s_4 < \Delta)$ and a hard gluon $(s_4 > \Delta)$ part \cite{svn}, where
$\Delta$  has to be much smaller than $s,t_1,u_1$ and $m^2$. The soft gluon
cross section, related to $\int_0^{\Delta} ds_4$, can then be obtained
analytically \cite{longpaper} and the singularities cancel 
in the sum with the virtual cross sections whereas the
hard gluon part contains only the left over M singularities which are
removed by factorization. 
 
As our final technical remark let us recall that the factorization 
`counter cross section' for the light quark initiated 
subprocess (\ref{eq:nloqua}) contains a part which can be 
schematically expressed as
\begin{equation}
\label{eq:fact}
d\Delta\sigma_{q\gamma}^{\mathrm{fact}} = -\frac{\alpha_s}{2\pi} \left[
\left( \frac{2}{n-4} \Delta P_{q\gamma} + \Delta F_{q\gamma} \right)
\otimes d \Delta\sigma_{q\bar{q}\rightarrow Q\bar{Q}} \right] + \ldots
\end{equation}
where $\otimes$ denotes a convolution, $\Delta P_{q\gamma}$ is the
usual LO spin-dependent photon-to-quark splitting function and
$\Delta F_{i\gamma}$ represents the freedom in choosing a factorization
prescription. In the $\overline{\mathrm{MS}}$ scheme (which we use) 
$\Delta F_{q\gamma}$ has the form
\begin{equation}
\label{eq:fact2}
\Delta F_{q\gamma}^{\overline{\mathrm{MS}}}(x,\mu_f,\mu) =
\Delta P_{q\gamma}(x) \left(\gamma_E - \ln 4\pi + 
\ln \frac{\mu_f^2}{\mu^2} \right)
\end{equation}
where $\gamma_E$ is the Euler constant,
$\mu$ is the mass scale introduced by $n$ dimensional regularization, 
and $\mu_f$ is the arbitrary factorization
scale (in other schemes one subtracts different terms in (\ref{eq:fact2})). 
By applying (\ref{eq:fact}) one introduces the parton content of the
real (on-shell) polarized {\em{photon}} \cite{photon} 
which is experimentally completely unknown for the time being. Strictly
speaking a physically consistent (i.e., scheme independent) result for
heavy quark photoproduction in ${\cal{O}}(\alpha_s^2\alpha)$ can only
be obtained for the sum of the `direct' {\em{and}} `resolved'
contribution due to the freedom in (\ref{eq:fact2}). In the first case
the photon acts as an elementary particle (as in the present calculation)
whereas in the latter case it resolves into its parton content before
the hard scattering takes place. The NLO corrections for the `resolved'
contribution are unknown so far for the polarized case and thus, for
the time being, have to be estimated in LO (see below).

%

Let us now turn to some numerical results and phenomenological
aspects. 
We are here only interested in the experimentally
most relevant total photoproduction cross section (for other results
see \cite{longpaper}). Let us start with the total photon-parton
cross section in NLO which can be expressed in terms of
scaling functions $(i=g,\,q,\,\bar{q})$
\begin{equation}
\Delta \hat{\sigma}_{i\gamma} (s,m^2,\mu_f)=
\frac{\alpha\alpha_s}{m^2} \left[ 
\Delta f_{i\gamma}^{(0)}(\eta) + 4\pi \alpha_s 
\left\{\Delta f_{i\gamma}^{(1)}(\eta) +
\Delta \bar{f}_{i\gamma}^{(1)}(\eta) \ln \frac{\mu_f^2}{m^2}\right\}
\right]
\label{eq:partonxsec}
\end{equation}
where $\Delta f_{i\gamma}^{(0)}$ and $\Delta f_{i\gamma}^{(1)}$,
$\Delta \bar{f}_{i\gamma}^{(1)}$ stand for the LO and NLO corrections,
respectively, $\mu_f$ denotes the factorization scale, 
and $\eta\equiv s/4m^2 -1$. Note that for simplicity 
we have made the conventional choice \cite{svn}
$\mu_r =\mu_f$ in (\ref{eq:partonxsec}) 
and in what follows (see \cite{longpaper} for an
independent variation of $\mu_f$ and $\mu_r$). 
The scaling functions can be further decomposed depending on the
electric charge of the heavy and light quarks, $e_Q$ and $e_q$,
respectively\footnote{Note that the interference term of the two
possible production mechanisms for the subprocess (\ref{eq:nloqua})
proportional to $e_Q e_q$ does not contribute to $\Delta f_{q\gamma}$
in (\ref{eq:fq}), since it vanishes when integrated over the
entire phase space as a consequence of the Furry theorem.}:
\begin{eqnarray}
\label{eq:fg}
\Delta f_{g\gamma} (\eta) &=& e_Q^2\, \Delta c_{g\gamma}(\eta)\\
\label{eq:fq}
\Delta f_{q\gamma} (\eta) &=& e_Q^2\, \Delta c_{q\gamma} (\eta) +
e_q^2\, \Delta d_{q\gamma} (\eta)
\end{eqnarray}
with corresponding expressions for the $\Delta \bar{f}_{i\gamma}$.

In Fig.~1 we present $\Delta c_{g\gamma}^{(0)}$, $\Delta c_{g\gamma}^{(1)}$,
and $\Delta \bar{c}_{g\gamma}^{(1)}$ as a function of the scaling
parameter $\eta$ in the $\overline{\mathrm{MS}}$ scheme (the threshold
$s=4m^2$ is located at $\eta=0$). 
Also shown (dotted line)
is the contribution to $\Delta c_{g\gamma}^{(1)}$ from the hard gluon
part $(s_4 > \Delta)$ alone\footnote{To define this quantity we follow the
procedure in \cite{svn} and explicitly add the $\Delta$ dependent terms
$(\sim \ln \Delta/m^2)$ of the soft plus virtual cross section to the hard
part in order to cancel the dependence on the auxiliary quantity 
$\Delta$ numerically.}.
From a comparison of this dotted curve with the dashed one
showing the full NLO coefficient $\Delta c_{g\gamma}^{(1)}$
(i.e., the sum of the hard and soft plus virtual contributions),
one can infer that the 
soft plus virtual contribution is almost negligible for $\eta\lesssim 1$.
For the interpretation of the results below it is important to notice that
the LO coefficient (solid line) in Fig.~1 changes sign at 
$\eta\simeq 3$. Upon adding the NLO contributions,
multiplied by a factor $4\pi \alpha_s$ (see Eq.~(\ref{eq:partonxsec})),
the zero is shifted towards $\eta\simeq 1$. 
We also note that for $\eta\lesssim 0.1$ the ${\cal{O}}(\alpha_s)$
corrections dominate over the Born approximation when we include that factor.
This enhancement near threshold is due to large logarithms from 
initial state gluon bremsstrahlung \cite{svn}. Furthermore in the
threshold limit $\eta\rightarrow 0$ the soft plus virtual piece tends
to a constant (`Coulomb singularity') while in LO 
$\Delta c_{g\gamma}^{(0)}\rightarrow 0$.

Fig.~2 compares in a similar way the light quark induced coefficients 
$\Delta c_{q\gamma}^{(1)}$, $\Delta \bar{c}_{q\gamma}^{(1)}$,
$\Delta d_{q\gamma}^{(1)}$, and $\Delta \bar{d}_{q\gamma}^{(1)}$ as
a function of $\eta$ in the $\overline{\mathrm{MS}}$ scheme.
Numerically they turn out to be much smaller than their
gluonic counterparts shown in Fig.~1.
Comparing our polarized and unpolarized results (for the latter see also
Figs.~5, 7, and 8 in \cite{svn}) for either the gluon or
the quark coefficients, we find the same pattern:
for $\eta \rightarrow 0$ the unpolarized and polarized results become equal, 
e.g., $\Delta c_{g\gamma}\rightarrow c_{g\gamma}$. A glance at
Eqs.~(\ref{eq:unpme}) and (\ref{eq:polme}) immediately implies that
$\left|M_{i\gamma}\right|^2(+-) \rightarrow 0$.
On the contrary, for asymptotically large energies 
$\eta\rightarrow \infty$ the unpolarized NLO coefficients
approach a large plateau value, except for $d_{q\gamma}^{(1)}$ and
$\bar{d}_{q\gamma}^{(1)}$, dominating over the LO result 
due to Feynman diagrams with a gluon exchange in the $t$-channel \cite{svn} 
while all polarized coefficients tend to zero in that kinematical region.
Thus one can infer that here 
$\left|M_{i\gamma}\right|^2(++)\rightarrow \left|M_{i\gamma}\right|^2(+-)$.

With the total partonic cross section in (\ref{eq:partonxsec}) at hand
is is now straightforward to calculate the total hadronic heavy flavor
photoproduction cross section via
\begin{equation}
\label{eq:xsec}
\Delta \sigma_{\gamma p}^Q(S_{\gamma p},m^2,\mu_f) = 
\sum_{f=q,\bar{q},g}\int\limits_{4m^2/S_{\gamma p}}^1 dx\, 
\Delta \hat{\sigma}_{f \gamma}(x S_{\gamma p},m^2,\mu_f)\,
\Delta f^p(x,\mu_f^2)  
\end{equation}
depending on the available photon-proton c.m.s.\ energy $S_{\gamma p}$.
Of course an expression similar to (\ref{eq:xsec}) also holds
for the unpolarized cross section $\sigma_{\gamma p}^Q$ with all
polarized quantities replaced by the corresponding unpolarized ones.
Instead of measuring $\Delta \sigma_{\gamma p}^Q$ in (\ref{eq:xsec})
directly (which requires the determination of the absolute
normalization), experiments will study the related
longitudinal spin asymmetry defined by
\begin{equation}
\label{eq:asym}
A_{\gamma p}^Q(S_{\gamma p},m^2,\mu_f) = 
\frac{\Delta \sigma_{\gamma p}^Q (S_{\gamma p},m^2,\mu_f)}
{\sigma_{\gamma p}^Q(S_{\gamma p},m^2,\mu_f)}\quad .
\end{equation}

In Fig.~3 we show the LO and NLO predictions for the total charm asymmetry
$A_{\gamma p}^c$, using $m=1.5\;\mathrm{GeV}$, $\mu_f=2m$, and
three different sets of polarized parton distributions \cite{grsv,gs} which
mainly differ in $\Delta g$, in the $\sqrt{S_{\gamma p}}$ region
accessible by the upcoming COMPASS experiment at CERN \cite{compass}.
Depending on the used muon beam energy (100 or 200 GeV) they will determine
$A_{\gamma p}^c$ for one average value of $\sqrt{S_{\gamma p}}$ of about
10 GeV. The NLO corrections in Fig.~3 turn out to be 
sizeable, depend strongly
on $\sqrt{S_{\gamma p}}$ and do not cancel in the ratio (\ref{eq:asym})
as one may naively expect. However, the physical origin of these large
corrections is readily explained in terms 
of the LO and NLO coefficient functions (see Fig.~1 and 2).

For $\sqrt{S_{\gamma p}}\simeq 10\,\mathrm{GeV}$ one probes $\eta$ values
from threshold up to $\eta\approx 10$. Note that in 
$\Delta \sigma_{\gamma p}^c$, see Eq.~(\ref{eq:xsec}), 
the coefficient functions for small values of $\eta$ 
(i.e., small $s=x S_{\gamma p}$) are convoluted
with the gluon distribution at the smallest possible $x$ 
$(x\ge 4m^2/S_{\gamma p})$ where $\Delta g$ is usually
large and vice versa. Since the NLO corrections to the gluonic
coefficient functions, which dominate for small $\eta$, decrease faster 
with increasing $\eta$ in the polarized case due to the zero, one should
expect the NLO asymmetry to be somewhat smaller than the LO one
in that particular $\sqrt{S_{\gamma p}}$ region {\em{if}} the
LO and NLO $\Delta g$ are not too different. Moreover,
the NLO shift of the zero towards smaller $\eta$ adds
contributions with opposite sign in the convolution already for smaller 
$S_{\gamma p}$. This should also lower the NLO asymmetry with respect to
the LO one for not too large $S_{\gamma p}$. Instead we find (see Fig.~3)
that the NLO asymmetries for GRSV \cite{grsv} and GS (A) \cite{gs} are larger 
than the LO ones for small $S_{\gamma p}$. But this is entirely due to the
badly constrained $\Delta g$ at large $x$ and thus should not be
taken too seriously: For 
$\sqrt{S_{\gamma p}}\simeq 10\,\mathrm{GeV}$ the convolution (\ref{eq:xsec})
samples $x\gtrsim 0.1$ and in both sets of polarized 
parton distributions \cite{grsv,gs}
the NLO $\Delta g$ is much larger than the LO one (especially in the case
of GS (A)). This is not the case for the unpolarized GRV distributions
\cite{grv} which we use to calculate $\sigma_{\gamma p}^c$ in 
(\ref{eq:asym}).
If we instead use NLO gluons for both the NLO {\em{and}} LO
asymmetries, we find the expected behaviour discussed above for GRSV
and GS (A). This is demonstrated for the GRSV set 
by the dotted curve in Fig.~3. 
The situation for GS (C) \cite{gs}, however, is even more complex since
in this case the gluon oscillates as well in the covered $x$ region.
Also obviously the {\em relative} NLO corrections (`$K$-factors') are always
very large around zeros in the LO asymmetry, as for the GS (C) curve
in Fig.~3 at around $\sqrt{S_{\gamma p}} \simeq 12\,\mathrm{GeV}$
(similarly for GS (A) and GRSV at larger 
$\sqrt{S_{\gamma p}} \simeq 30\,\mathrm{GeV}$ not shown in the
figure), but this is natural for quantities that can change sign.

Let us finally briefly discuss the importance of the main theoretical 
uncertainties in the calculation of $\Delta \sigma_{\gamma p}^c$ or
$A_{\gamma p}^c$ which may further complicate the extraction of $\Delta g$.
First of all Fig.~4 shows the dependence of 
$\Delta \sigma_{\gamma p}^c$ on the
choice of the factorization scale in the relevant region
$\mu_f^2 =[m^2,\ldots,4m^2]$ for the GRSV standard set of polarized parton
densities \cite{grsv} for two different values of $\sqrt{S_{\gamma p}}$.
The NLO results hardly depend on the precise choice of $\mu_f$ 
whereas the LO cross sections vary
by as much as $50\%$ in the shown range. This clearly underlines the
importance (and usefulness) of the full NLO calculation. NLO results
for $\Delta \sigma_{\gamma p}^c$ or $A_{\gamma p}^c$ are thus much more
trustworthy than previous LO estimates.

As already mentioned above another complication in the extraction
of $\Delta g$ arises due to the presence of the light quark induced
subprocesses (\ref{eq:nloqua}) which may serve as an important 
`background'. We have checked that for 
$\sqrt{S_{\gamma p}} \simeq 10\,\mathrm{GeV}$ this contribution is
fairly small (about $5\%$, except for the GS (C) set 
where the gluon contribution is close to zero at this
$\sqrt{S_{\gamma p}}$) and can possibly be neglected 
for a first
determination of $\Delta g$. However, with a better knowledge of the 
polarized quark densities in the future the light quark contribution
should be subtracted.
A similar remark holds for the importance of the possible `background'
from `resolved' photons. Since the parton content of polarized photons
is completely unknown so far, one has to impose some realistic models
\cite{photon} to estimate this contribution. In \cite{svhera} it was 
shown that even for rather large photonic densities this `background' should
also be very small in the `COMPASS region'.
Moreover, the unknown precise 
value of the charm quark mass $m$ leads to shifts
in $\Delta \sigma_{\gamma p}$ and $A_{\gamma p}^c$ of about
$30\%$ when $m$ is varied by 0.2 GeV
around the central values of 1.5 GeV used in Figs.~3 and 4.

Finally, it should be stressed that for a meaningful and reliable
extraction of $\Delta g$ from a measurement of $A_{\gamma p}^c$
(i.e., roughly a measurement of $\Delta g/g$) our knowledge of the
{\em{unpolarized}} gluon density has to be improved as well.
At $\sqrt{S_{\gamma p}}\simeq 10\,\mathrm{GeV}$
one probes the gluon at $x$ values larger than 0.1
where the uncertainty in $g(x,\mu^2)$ is sizeable and non-negligible
(see, e.g., \cite{cteqg,mrst}). In that particular $x$ region
$g(x,\mu^2)$ is mainly constrained by direct photon data, but
with rather large theoretical uncertainties \cite{vv}.
Ideally, COMPASS should try to measure also the unpolarized charm
photoproduction cross section, thereby reducing also our present ignorance on
$g$ at large $x$. Anyway, improvements seem to be mandatory for a reliable
extraction of $\Delta g$.

%

To summarize, we have presented the main results of a first complete
NLO QCD calculation for heavy flavor photoproduction with 
longitudinally polarized beams. The NLO corrections are sizeable for the
total longitudinal spin asymmetry in the energy range accessible by the
forthcoming COMPASS experiment. But these corrections 
can be partly attributed to the rather different large 
$x$ behaviour in LO and NLO of the presently 
available sets of polarized parton densities, which only reflects our 
complete ignorance of $\Delta g$.
The theoretical uncertainties associated with the precise choice 
of a factorization scale were shown to be strongly
reduced in NLO. The prospects for a first direct measurement
of the polarized gluon density still seem to be promising but a
careful and reliable extraction of $\Delta g$ requires also an
improved knowledge of the unpolarized gluon density at large $x$
and of the precise value of the charm quark mass. Uncertainties
due to the `background' from `resolved' photons or light quark induced
processes should to be of less importance.

\section*{Acknowledgments}
We are grateful to W.\ Vogelsang for useful discussions. I.B. wishes
to thank E.\ Reya for suggesting this problem and 
his constant encouragement, and W.\ Beenakker
for helpful comments on some integrals. 
This work has been supported in part by the
`Bundesministerium f\"ur Bildung, Wissenschaft, Forschung und
Technologie', Bonn.
\newpage
%

%
\newpage
%
\section*{Figure Captions}
\begin{description}
\item[Fig.\ 1] The LO and NLO gluonic scaling functions in the 
$\overline{\mathrm{MS}}$ scheme as a function of $\eta=s/4m^2-1$ as
defined in Eqs.~(\ref{eq:partonxsec}) - (\ref{eq:fq}). 
\item[Fig.\ 2] Same as in Fig.~1, but for the NLO light quark coefficient
functions.
\item[Fig.\ 3] The longitudinal spin asymmetry for total charm quark
photoproduction in LO and NLO as defined in (\ref{eq:asym}) for
$m=1.5\,\mathrm{GeV}$, $\mu_f=2m$ and three different sets of
LO and NLO polarized parton densities \cite{grsv,gs}. The unpolarized
cross section in (\ref{eq:asym}) was calculated using the GRV \cite{grv}
densities. Also shown (dotted line) is the LO asymmetry using the NLO
GRSV \cite{grsv} gluon distribution (see text).
The vertical bar shows the estimated statistical uncertainty
$\delta A_{\gamma p}^c$ for such a measurement at COMPASS \cite{compass}.  
\item[Fig.\ 4] The factorization scale dependence for the LO and NLO
polarized total charm quark photoproduction cross section 
$\Delta \sigma_{\gamma p}^c$ as defined in
(\ref{eq:xsec}) using the LO and NLO GRSV standard densities \cite{grsv},
respectively, for $m=1.5\,\mathrm{GeV}$.
\end{description}
\newpage
%
\pagestyle{empty}
\begin{center}

\vspace*{-1.0cm}
\hspace*{-1.cm}
\epsfig{file=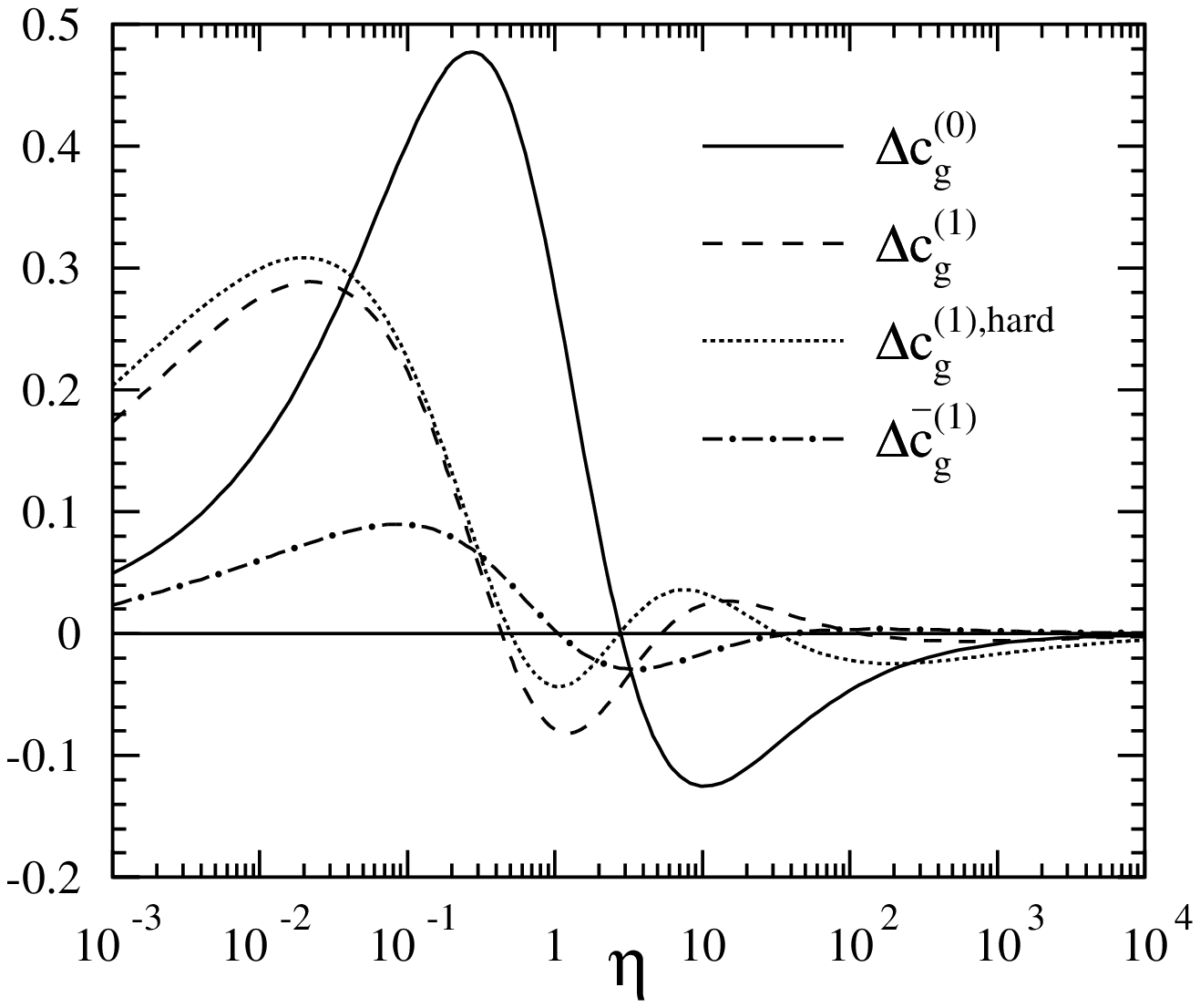}

\vspace*{-0.8cm}
\Large{\bf{Fig.\ 1}}
\end{center}

\begin{center}

\vspace*{-1.0cm}
\hspace*{-1.cm}
\epsfig{file=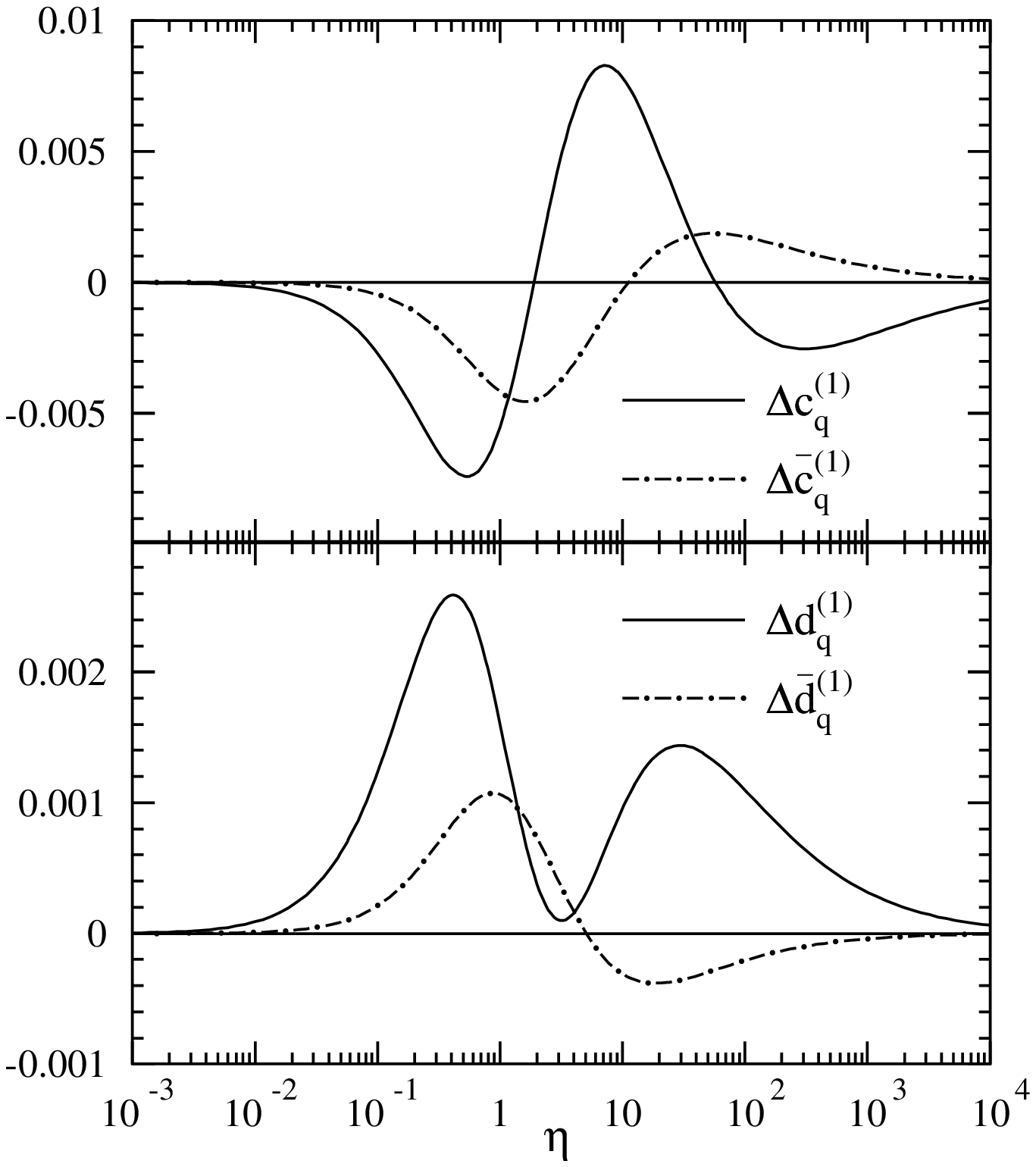}

\vspace*{-0.8cm}
\Large{\bf{Fig.\ 2}}
\end{center}

\begin{center}

\vspace*{-1.0cm}
\hspace*{-1.cm}
\epsfig{file=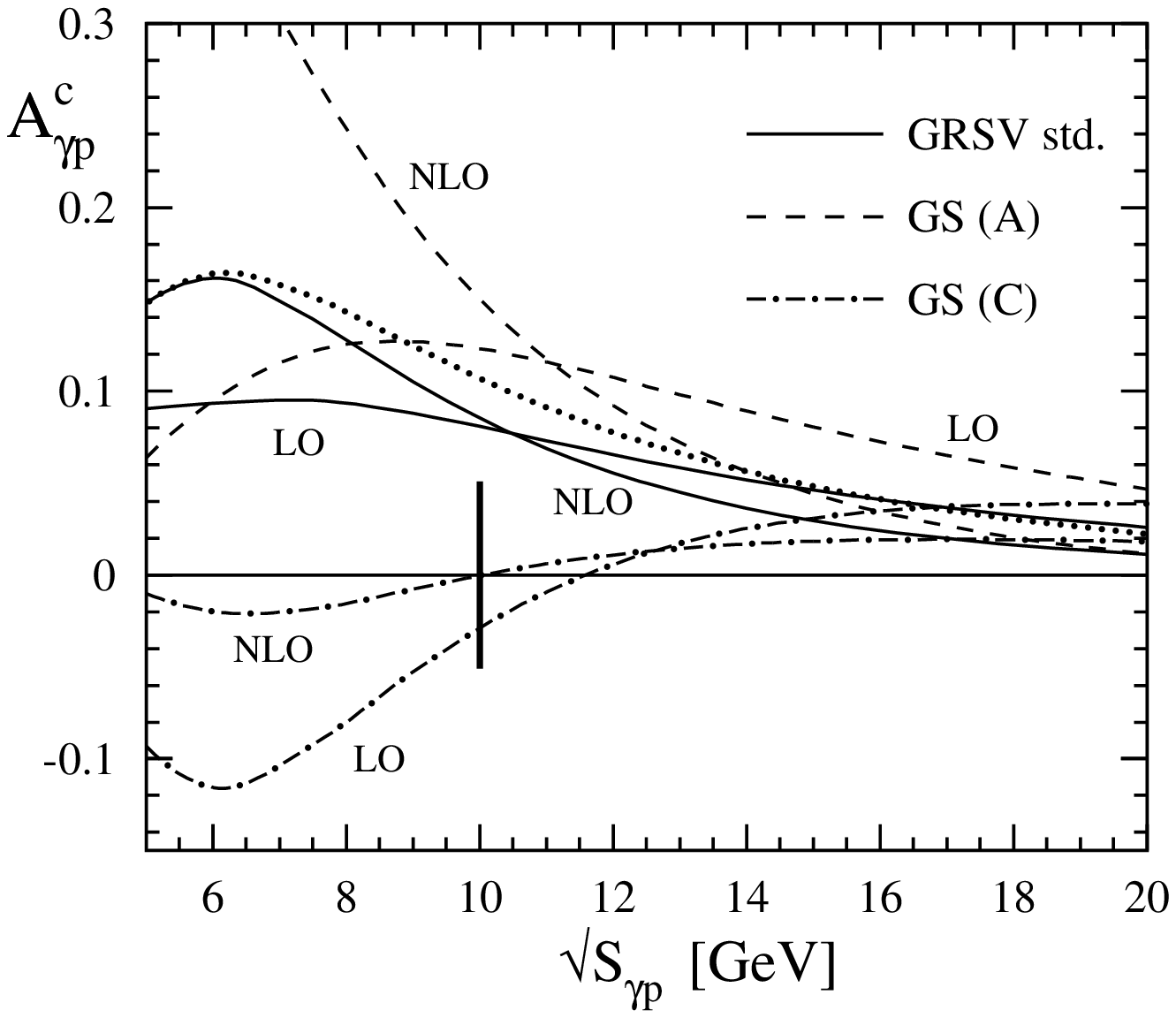}

\vspace*{-0.8cm}
\Large{\bf{Fig.\ 3}}
\end{center}

\begin{center}

\vspace*{-1.0cm}
\hspace*{-1.cm}
\epsfig{file=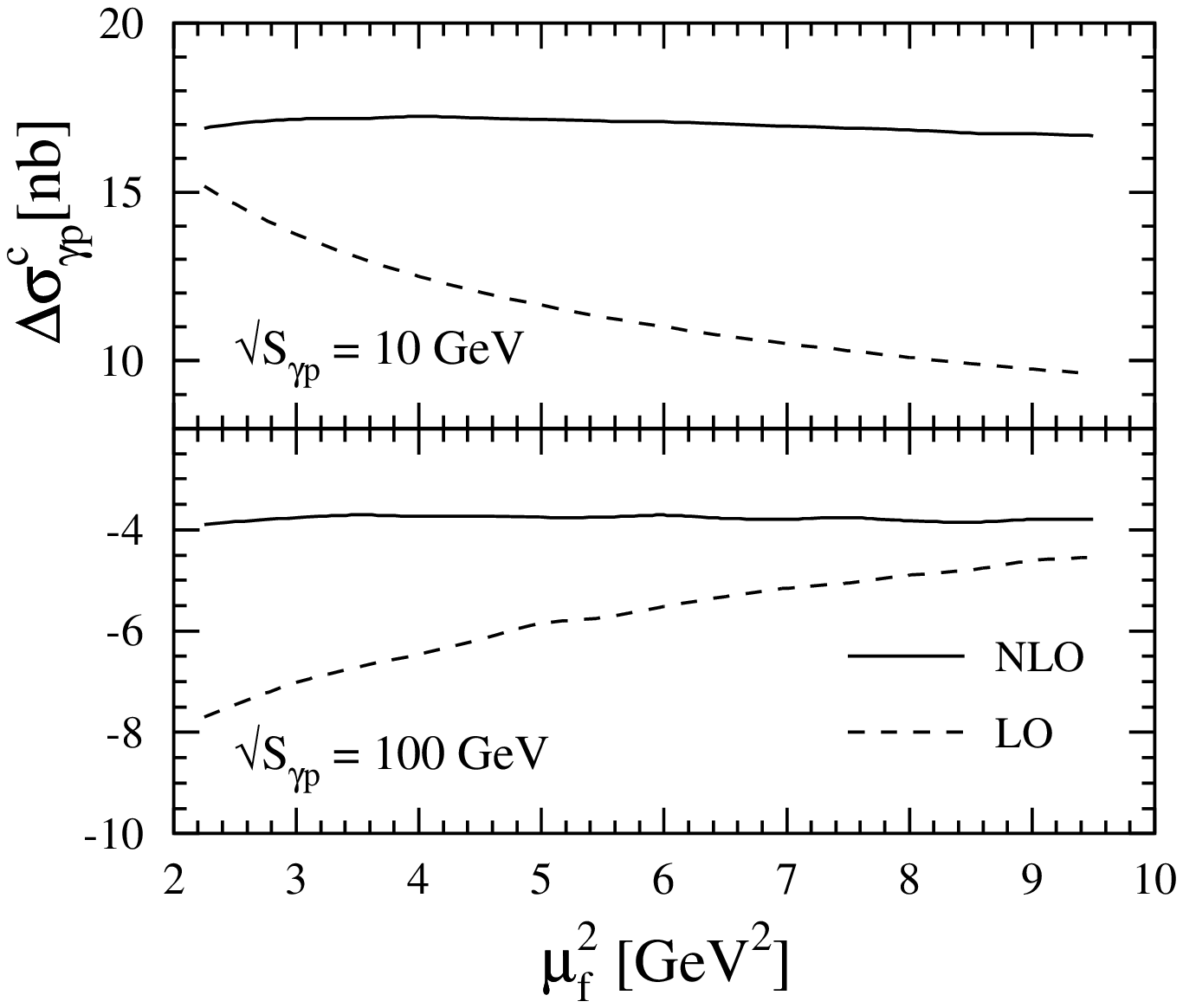}

\vspace*{-0.8cm}
\Large{\bf{Fig.\ 4}}
\end{center}

\begin{thebibliography}{99}
%
%
\bibitem{exp} A recent overview of the experimental status can be found
for example in the proceedings of the workshop `Deep Inelastic Scattering
off Polarized Targets: Theory Meets Experiment', Zeuthen, Germany, 1997,
ed.\ by J.\ Bl\"umlein et al., DESY 97-200, pp.\ 2-40.
%
\bibitem{grsv} M.\ Gl\"uck, E.\ Reya, M.\ Stratmann, and W.\ Vogelsang, 
Phys. Rev. {\bf D53}, 4775 (1996).
%
\bibitem{gs} T.\ Gehrmann and W.J.\ Stirling, Phys. Rev. {\bf D53},
6100 (1996). 
%
\bibitem{pdfs} G.\ Altarelli, R.D.\ Ball, S.\ Forte, and G.\ Ridolfi,
Nucl. Phys. {\bf B496}, 337 (1997), CERN-TH-98-61, {\tt hep-ph/9803237};\\
D.\ de Florian, O.\ Sampayo, and R.\ Sassot, Phys. Rev. {\bf D57}, 
5803 (1998);\\
E.\ Leader, A.V.\ Sidorov, and D.B.\ Stamenov, {\tt hep-ph/9708335};\\
K.\ Abe et al., E154 collab., Phys. Lett. {\bf B405}, 180 (1997);\\
D.\ Adams et al., SMC, Phys. Rev. {\bf D56}, 5330 (1997).
%
\bibitem{compass} G.\ Baum et al., COMPASS collab., 
CERN/SPSLC-96-14, CERN/SPSLC-96-30.
%
\bibitem{lohq} M.\ Gl\"uck and E.\ Reya, Z. Phys. {\bf C39}, 569
(1988).
%
\bibitem{svhera}  M.\ Stratmann and W.\ Vogelsang, Z. Phys. {\bf C74},
641 (1997). 
%
\bibitem{other} G.\ Altarelli and W.J.\ Stirling, Particle World {\bf 1},
40 (1989);\\
M.\ Gl\"uck, E.\ Reya, and W.\ Vogelsang, Nucl. Phys. 
{\bf B351}, 579 (1991);\\
S.I.\ Alekhin, V.I.\ Borodulin, and S.F.\ Sultanov, Int. J. Mod. Phys.
{\bf A8}, 1603 (1993);\\
S.\ Keller and J.F.\ Owens, Phys. Rev. {\bf D49}, 1199 (1994);\\
S.\ Frixione and G.\ Ridolfi, Phys. Lett. {\bf B383}, 227 (1996).
%
\bibitem{svn} J.\ Smith and W.L.\ van Neerven, Nucl. Phys. {\bf B374},
36 (1992).  
%
\bibitem{ellis} R.K.\ Ellis and P.\ Nason, Nucl. Phys. {\bf B312},
551 (1989).
%
\bibitem{longpaper} I.\ Bojak and M.\ Stratmann, in preparation.
%
\bibitem{nason} P.\ Nason, S.\ Dawson, and R.K.\ Ellis, Nucl. Phys.
{\bf B303}, 607 (1988), {\bf B327}, 49 (1989).
%
\bibitem{craigie} N.S.\ Craigie, K.\ Hidaka, M.\ Jacob, and F.M.\
Renard, Phys. Rep. {\bf 99}, 69 (1983). 
%
\bibitem{conto} B.\ Kamal, Z.\ Merebashvili, and A.P.\ Contogouris,
Phys. Rev. {\bf D51}, 4808 (1995), {\bf D55}, 3229(E) (1997);\\ 
G.\ Jikia and A.\ Tkabladze, Phys. Rev. {\bf D54}, 2030 (1996). 
%
\bibitem{hvbm} G.\ t'Hooft and M.\ Veltman, Nucl. Phys. {\bf B44}, 189 
(1972);\\ P.\ Breitenlohner and D.\ Maison, Comm. Math. Phys. {\bf 52},
11 (1977).
%
\bibitem{tracer} M.\ Jamin and M.E.\ Lautenbacher, Comput. Phys. Commun.
{\bf 74}, 265 (1993). 
%
\bibitem{smith2} W.\ Beenakker, H.\ Kuijf, W.L.\ van Neerven, and J.\
Smith, Phys. Rev. {\bf D40}, 54 (1989).
%
\bibitem{werner} L.E.\ Gordon and W.\ Vogelsang, Phys. Rev. {\bf D48}, 
3136 (1993). 
%
\bibitem{pass} G.\ Passarino and W.\ Veltman, Nucl. Phys. {\bf B160},
151 (1979);\\
W.J.P.\ Beenakker, Ph.D.\ thesis, Univ.\ Leiden.
%
\bibitem{photon} M.\ Gl\"uck and W.\ Vogelsang, Z. Phys. {\bf C55},
353 (1992), {\bf C57}, 309 (1993);\\
M.\ Gl\"uck, M.\ Stratmann, and W.\ Vogelsang,
Phys. Lett. {\bf B337}, 373 (1994);\\ 
M.\ Stratmann and W.\ Vogelsang, Phys. Lett. {\bf B386}, 370 (1996).
%
\bibitem{grv} M.\ Gl\"uck, E.\ Reya, and A.\ Vogt, Z. Phys. {\bf C67}, 
433 (1995).
%
\bibitem{cteqg} J.\ Huston et al., CTEQ collab., 
FERMILAB-PUB-98-046-T, {\tt hep-ph/9801444}.
%
\bibitem{mrst} A.D.\ Martin, R.G.\ Roberts, W.J.\ Stirling, and R.S.\
Thorne, {\tt hep-ph/9803445}.
%
\bibitem{vv} W.\ Vogelsang and A.\ Vogt, Nucl. Phys. {\bf B453}, 334 (1995). 
%
\end{thebibliography}
\end{document}